\documentclass[aps,prd,onecolumn,12pt,preprintnumbers]{revtex4-1}

\usepackage{hyperref}

\usepackage{float}

\usepackage{tikz}

\usepackage{graphicx}
\usepackage{epstopdf}

\begin{document}

\preprint{HDP: 16 -- 03}

\title{Banjo Rim Height and Sound in the Pot}

\author{David Politzer}

%\email[]{politzer@theory.caltech.edu}
\email[]{politzer@theory.caltech.edu}

\homepage[]{http://www.its.caltech.edu/~politzer}

%\thanks{452-48 Caltech, Pasadena CA 91125}
\altaffiliation{\footnotesize Pasadena CA 91125}
%\altaffiliation{\newline \em \em \em 452-48 Caltech, Pasadena CA 91125}
\affiliation{}

\date{\today}
%\date{July 31, 2016}

\begin{figure}[h!]
\includegraphics[width=4.0in]{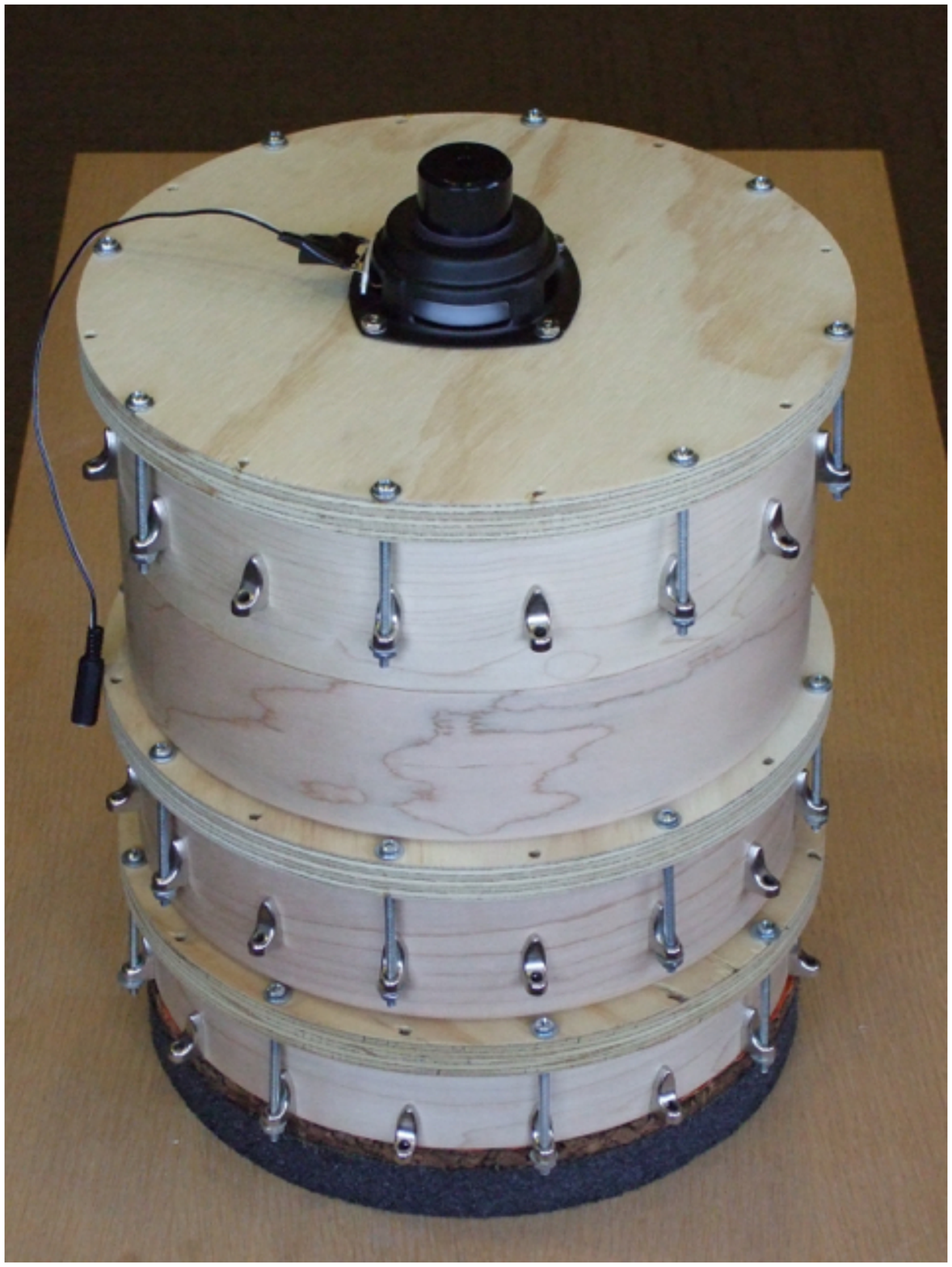}
\end{figure}

\begin{abstract}
Rim and back geometry determine much of the behavior of sound inside the pot, whose effect on total, produced sound is subtle but discernible.  The theory of sound inside a cylinder is reviewed and demonstrated.  And previous work on the Helmholtz resonance and the interplay between the Helmholtz resonance and the lowest head mode is revisited using some improved techniques. 

\end{abstract}

\maketitle{ {\centerline{\large \bf Banjo Rim Height and Sound in the Pot}}

\bigskip

\bigskip

\section{introduction \& outline}

Some years ago, Joe Dickey offered a simple physics model of the banjo.\cite{dickey}  He considered ideal strings attached through a point mass to the center of an ideal drum head.   With enough approximations and simplifications, such strings and drum head are soluble systems.  And the model allows one to follow the action from an initial string pluck to the radiated sound.  Another relevant system is the air motion inside the pot.  With some simple approximations, it is also, by itself, a soluble system.  However, there are two caveats.  While its impact is relevant to the concerns of builders and players, it is admittedly only a small piece of the total.  And, perhaps more significantly, the coupling of the inside air to the head is strong but not well understood.  Internal air pressure variations make contact over the entire surface of the head, and that certainly effects how the head moves.  But how that plays out has not been studied in any particular detail.

\begin{figure}[h!]
 \includegraphics[width=5.0in]{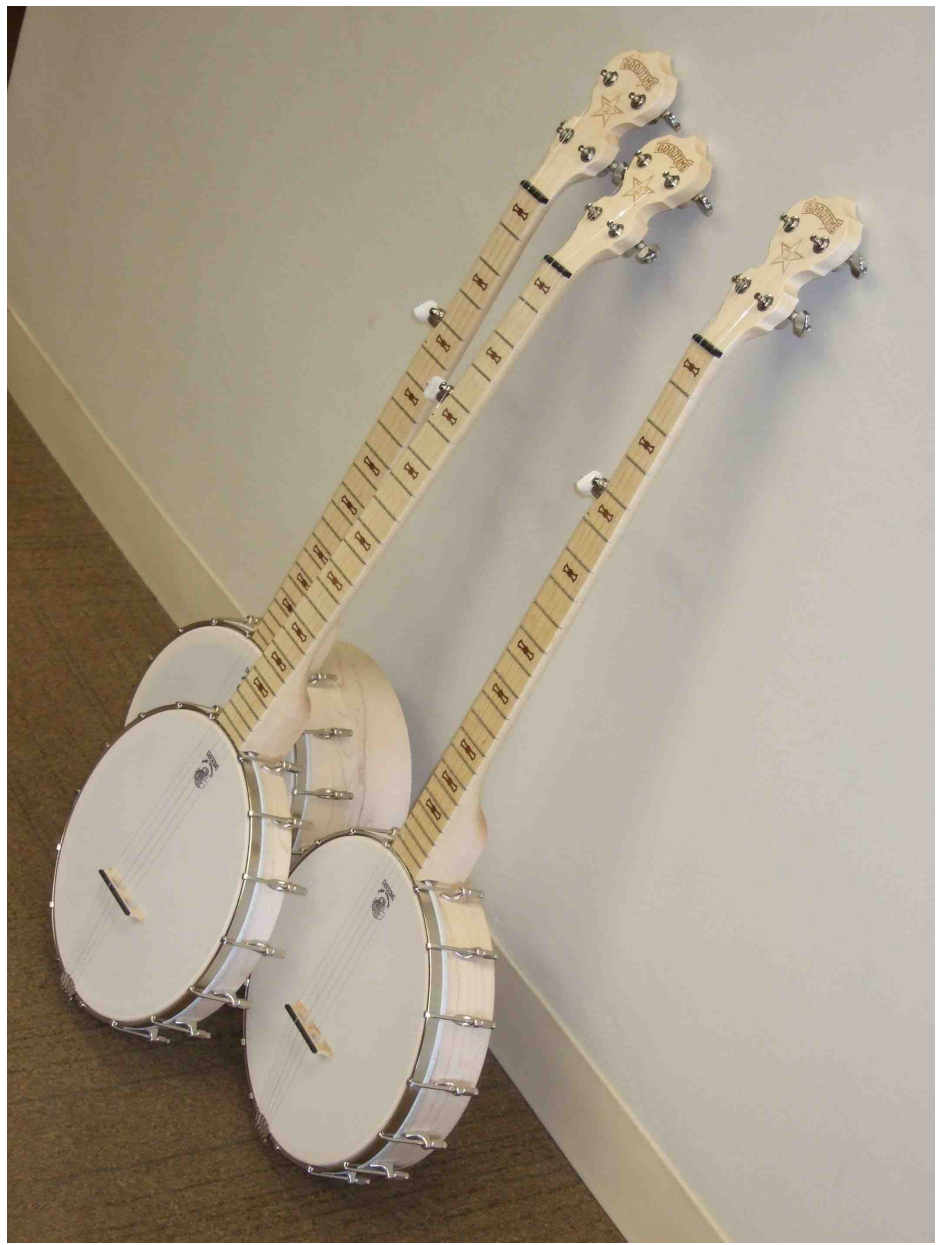}
 \caption{shallow, standard, \& deep}
 \end{figure}

This note is really just an addendum to an earlier work, {\it The Open Back of the Open-Back Banjo}.\cite{openback}  That was an investigation of the effects of rim height on air loading of the lowest frequency head motion and on the pot's Helmholtz resonance.  Here I compare the well-understood calculation of sound resonances of cylindrical cavities to measurements on those same three banjos, identical except for their rim heights (shown in FIG.~1).  Again, admittedly, rim height relevance is somewhat indirect.  The banjo's sound comes overwhelmingly from the vibration of the head.  The influence of the pot internal sound is through its coupling to the head.  That interaction is understood only qualitatively rather than in detail.

Another caveat concerns the significant differences between the transient response due to a pluck and the steady-state response to continuous driving.  Dickey's strings and head and virtually all discussions in the acoustics of musical instruments consider systems in terms of their steady-state response.  Transients of coupled systems, even if they are linear, are more complex.\cite{coupled-damped}  (For example, the modes of specific frequencies are, in general, not normal or orthogonal.)  Nevertheless, for systems where the damping is weak, the steady-state resonant spectrum is a good starting point.

In the following, I give a verbal description of the sound resonances of cylindrical cavities.  Rim height is identified as a crucial variable in determining the qualitative behavior of the spectrum.  For the Helmholtz resonance, an essential feature is the air going in and out of the cavity.  For the other cavity resonances, it is a huge simplification and a realistic approximation to consider and measure a closed volume.  Sound spectrum measurements with three different rims are compared with each other and with the simple physics predictions.  The previous Helmholtz resonance and head air-loading measurements\cite{openback} are repeated.  I attempt to explain the shortcomings of a ballistic picture of sound propagation.  

\section{cylindrical cavity air resonances}

For sound waves (unless we're concerned with very narrow apertures), pressure is the only relevant force.  The equations of motion are standard undergraduate physics fare, and the cylinder solutions are among the simplest of three dimensional examples, after the rectangular box.  The relevant cylindrical coordinates are shown in FIG.~2.  
\begin{figure}[h!]
 \includegraphics[width=1.8in]{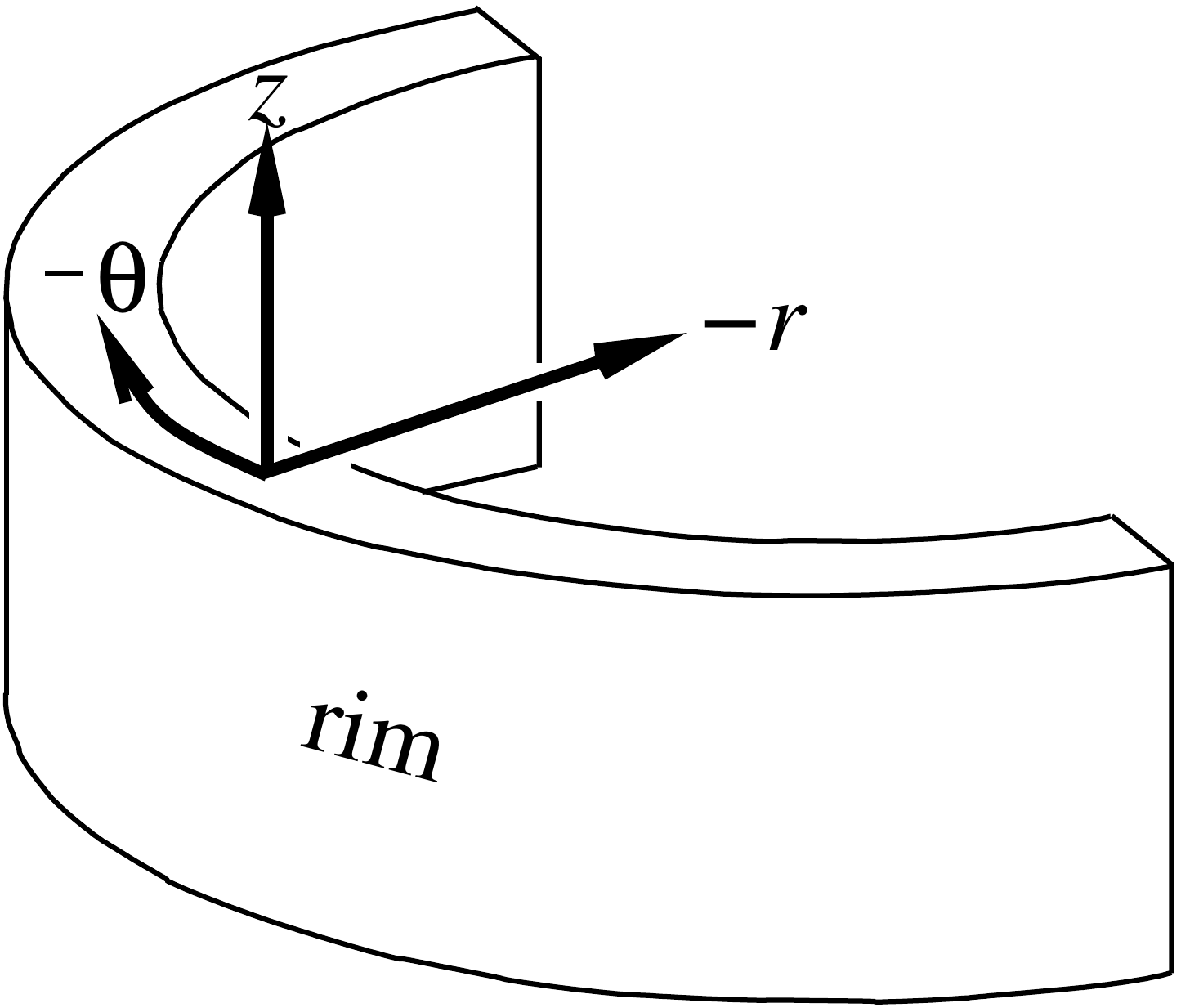}
 \caption{defining directions and coordinates relative to the rim}
 \end{figure}

The air pressure resonance solutions have specific frequencies and are products of a function of $r$ times a function of $\theta$ times a function of $z$.   The $z$-dependence is the simplest and the most relevant to the question of rim height effects.  The $z$ pressure function is sinusoidal, with maxima at the top and bottom of the cylinder.  So this is a series of integer numbers of half waves that fit in the cylinder.  Importantly, the series starts with zero.  The lowest frequency $z$ contribution to the total pressure function is independent of $z$.  So, for a squat cylinder, the several lowest resonances have oscillating pressures that are independent of $z$.  The air motion at those resonant frequencies is purely in the $r$-$\theta$ plane.  That also means that cylinders with the same diameter have the same resonant frequencies, independent of their $z$ total dimension (rim height), at least until the first $z$-dependent resonance is reached.

The lowest $z$-dependent resonance is independent of $r$ and $\theta$ and is simply a single half wave.  So its wavelength is just twice the rim height, and its period is the time it takes for sound to bounce once back and forth from top to bottom to top.  And the frequency is one over the period.

It is particularly noteworthy how the $r$, $\theta$, and $z$ motions combine to produce resonant frequencies above that lowest $z$-dependent resonance.  In particular, for the combined motions, the frequencies of the separate factors ``add in quadrature.," i.e., we take the square root of the sum of the squares --- like finding the hypotenuse of a right triangle.  In particular, there is a series of frequencies that involve motion in the $r$-$\theta$ plane that are independent of rim height.  Above the first $z$-dependent resonance, we multiply that $z$-dependence with the series of $r$-$\theta$ resonance functions to get the total pressure dependence.  The frequency of the product is the two separate frequencies added in quadrature.  Again for the squat cylinder, there are many $z$-independent resonances before the first $z$ mode appears.  When that mode is ``dressed" with the possible $r$ and $\theta$ dependences, the resulting sequence is much closer spaced in frequency than the original $z$-independent series --- at least in cases where the lowest $z$ frequency is  much higher than the $r$-$\theta$ frequencies in question.

\begin{figure}[h!]
 \includegraphics[width=5.0in]{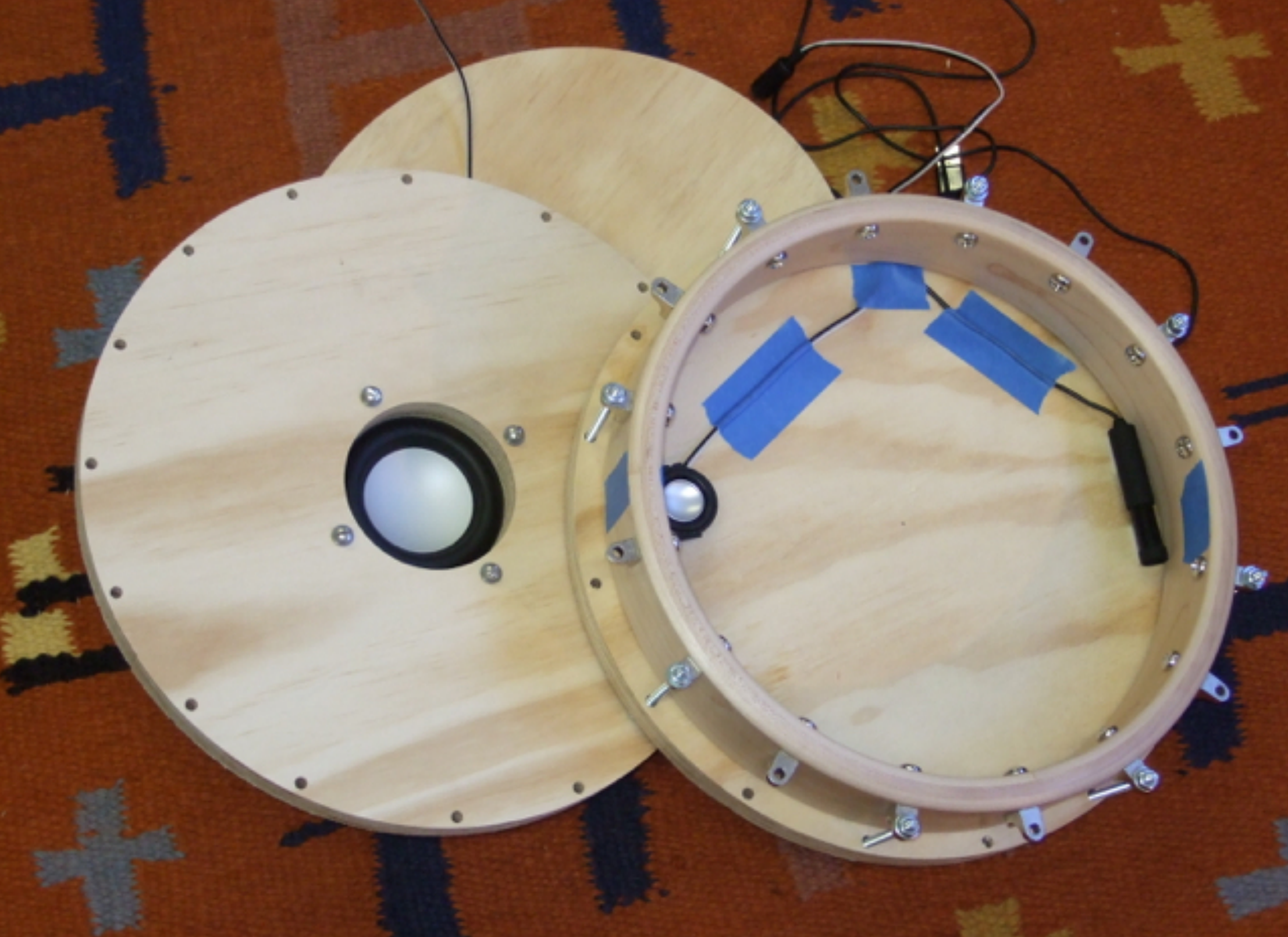}
 \caption{Speaker and microphone mounted internally on a solid plywood head and a second plywood disk to close the back; also the head-speaker combination used for FIG.s~6 and 9}
 \end{figure}

Measurements were carried out on three Goodtime rims.\cite{Goodtime}  Their heights were $2.04''$, $2.80''$, and $5.69''$.\cite{heights}  Acoustical investigations of guitars and violins have sometimes gone to great lengths to decouple the soundboard, side, and back vibrations from the vibrations of the air inside.  For example, the whole instrument might be buried in sand.  For a banjo, it's much easier.  I simply replaced the head with $3/4''$ plywood, and attached another plywood disk to the back.  The sound was excited by a $3/4''$ speaker mounted inside and recorded by a small microphone, also mounted inside.  The speaker and microphone were diametrically opposite, just under the head, as shown in FIG.~3.  (Their locations and physical extent limit what resonances can be detected; for example, you cannot detect a mode if either speaker or mic lie on a node line.)  I used a signal generator and audio amplifier to sweep linearly in frequency from 200 to 4000 Hz.  And the signal was recorded and analyzed using Audacity{\footnotesize$^{\textregistered}$}.

\begin{figure}[h!]
 \includegraphics[width=6.5in]{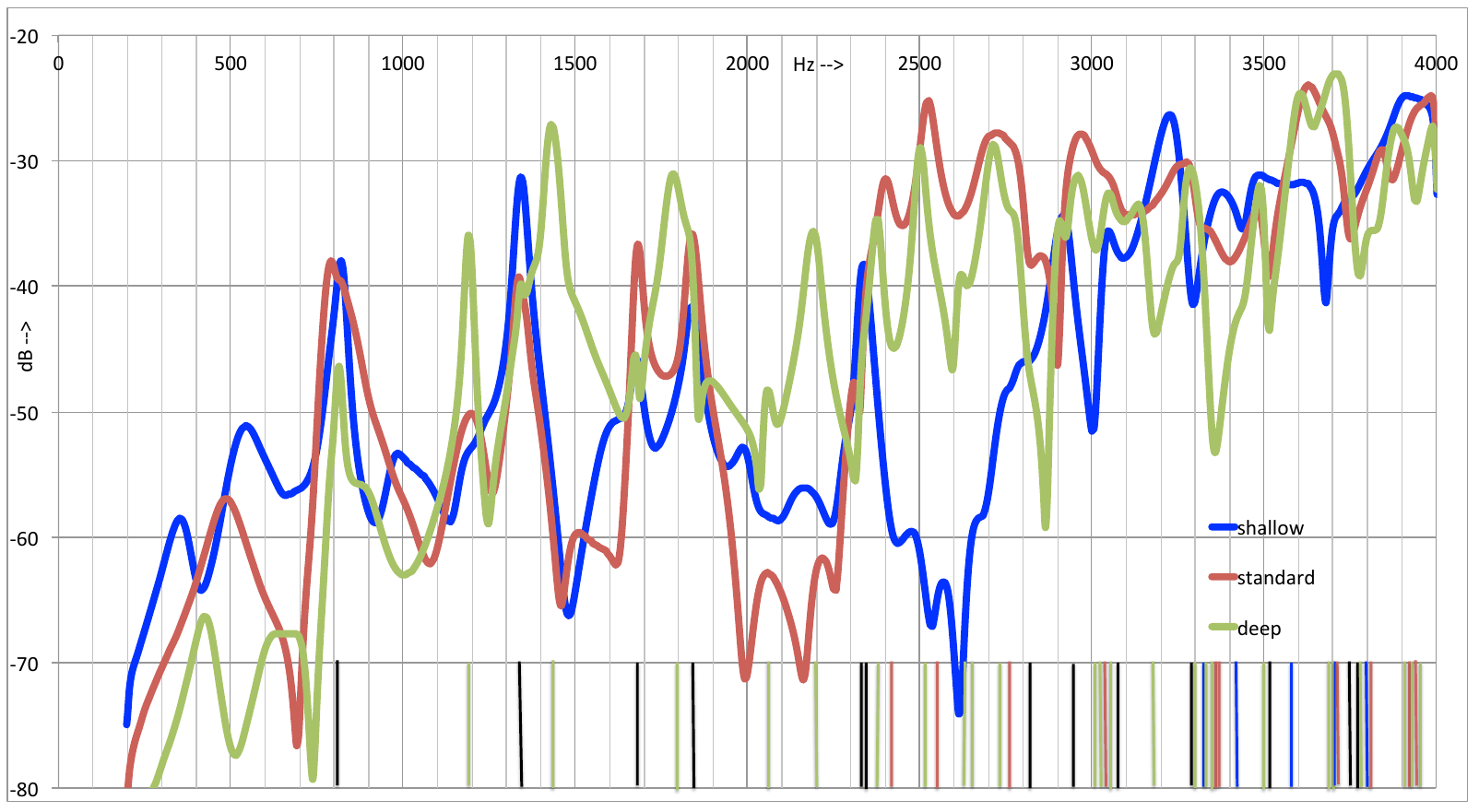}
 \caption{Measured spectra of the three rims; lines at the bottom are values computed from the diameters and heights; the black lines are the resonances, common to all, that have no variation in the $z$ direction. }
 \end{figure}

In FIG.~4, curves for the three rims are labeled shallow, standard, and deep for the three rim heights, respectively.  The straight lines at the bottom of the graph indicate the calculated values of the resonant frequencies.  The black lines are the $z$-independent $r$-$\theta$ modes for cylinders of internal diameter $9.85''$.  (The three Goodtime inner diameters were within 0.6\% of each other.)  For each rim, there is a lowest frequency corresponding to a half-wave in $z$.  Those frequencies are calculated to be 3322 Hz, 2421 Hz, and 1192 Hz, for shallow, standard, and deep, respectively.  Those and the additional $r$-$\theta$ frequencies combined in quadrature give the sequences of color-coded lines.  Note that under 4000 Hz, the deep pot is the only one which has a second series starting at the $z$ mode corresponding to two half wavelengths within the pot.  That one starts at 2384 Hz.

There are at least three noteworthy features, which I'll discuss before addressing some caveats and limitations below.  First, the calculated values match the tall peaks in the measurement.  Impressive or not, that's what physics is supposed to do.  Second, for each rim, at the lowest $z$ resonance and above, the resonances are closer together than in the absence of that $z$ mode.  And, of  course, its location is a simple function of rim height.  These resonances interact with the head and influence how well the head can convert string vibrations into sound.  So the closer internal cylinder resonances should give a more even response.  And third and somewhat more subtle, there is a step up in overall response (likely related to the resonances being closer together in frequency) above the first $z$ resonance.  In fact, the deep pot exhibits a second step. 

The lowest calculated resonance is at 808 Hz, and, indeed, all three pots show a fine peak very close to that value.  The stuff below 808 Hz (albeit not vey loud) is a clear indication that there are other things vibrating besides the air in the cylinder.  In fact, the observations between 200 and 800 Hz varied from run to run, but  I ran out of patience trying to track down every origin of the variability.  The heavy plywood and firm bolting of the head were likely not the culprits.  But the double-sticky foam tape and masking tape mounting of the speakers, microphones, and wires might not be as reproducible as some properly machined apparatus.  Similar small variability was also observed across the whole frequency range, likely of the same origin.  However, the prominent, high peaks were always identifiable at nearly the same frequencies.

The strength of a resonance in a system of multiple parts depends on two things.  First, how well does the driving match the geometry of the resonance?  To get a strong response, you have to push in the right place.  Pushing in the wrong way might not get any response at all.  And second, the effectiveness of a push of a fixed frequency depends on how close that frequency is to the resonant frequency.  Both of these must be kept in mind when addressing the real question of interest of does string vibration turn into sound.

\section{helmholtz and head resonances}

The theory here is crude but simple.  For pots that differ only by their rim height, the Helmholtz frequency should be inversely proportional to the square root of the height.  The lowest head mode couples strongly to the Helmholtz mode. In the absence of that coupling, its frequency would be inversely proportional to the square root of an increasing, linear function of the height.  (See ref.~\cite{openback}.)

The following is an attempt to identify the lowest frequency head modes.  All three heads  were set to the same tension as determined by a DrumDial (at 89).  The heads were tapped at their centers with a piano hammer.  The sound of a long series of taps was recorded with a microphone at $12''$ in front of the center of the head.  The (open) backs were left wide open.  Hence, there is substantially less air springiness than would be provided by a more enclosed volume.  Increased pot depth increases the ``air loading."  That is usually thought of as the air that the head has to move if it moves.  At a minimum, the effect is to increase the effective inertia or mass of the head.  Because heads are very thin, this is a much bigger deal than it is for the soundboard of wood-topped instruments.  An effect in the expected direction is evident in FIG.~5.  However, interpreting these peak locations in the context of the head interaction with the Helmholtz resonance will prove problematic.
\begin{figure}[h!]
 \includegraphics[width=6.5in]{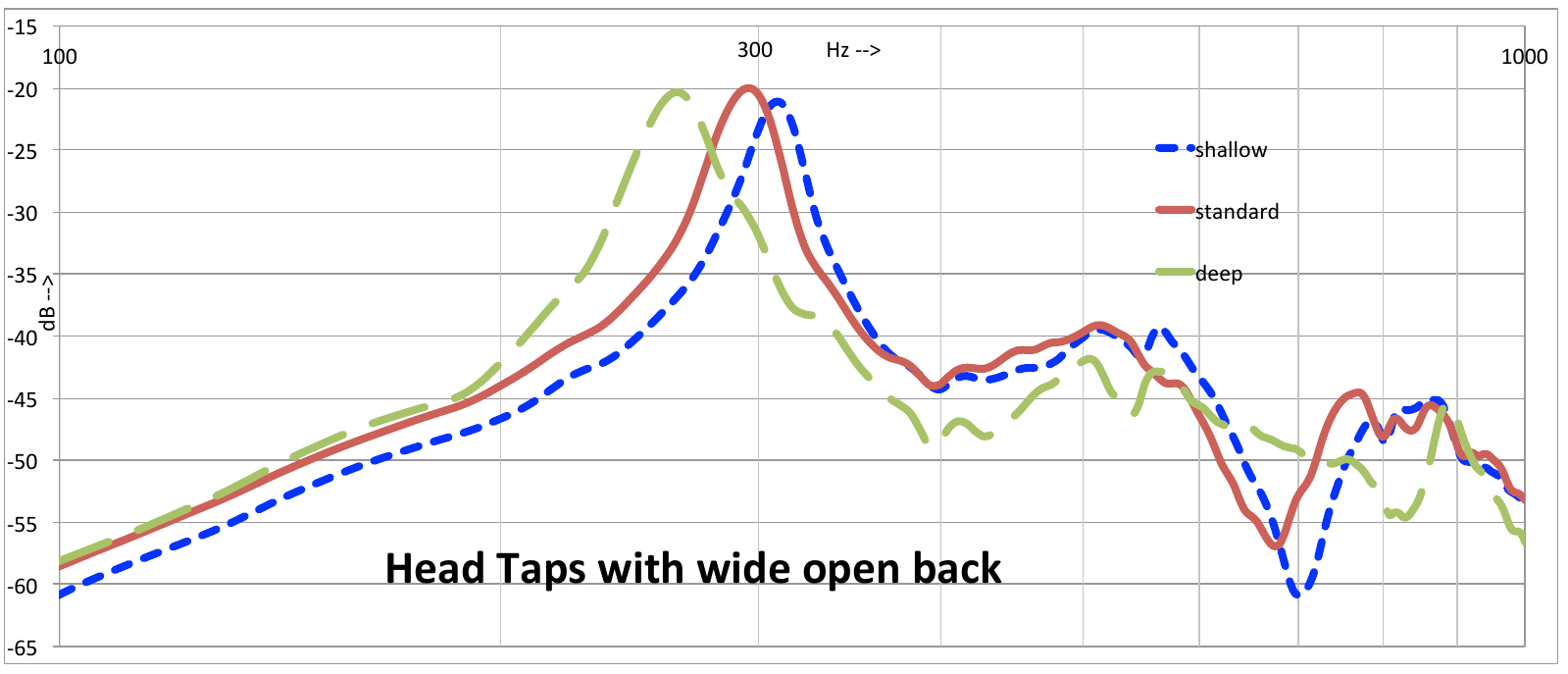}
 \caption{The sound of head taps with wide open backs, recorded in front }
 \end{figure}

The Helmholtz resonances can be decoupled from the head by using a plywood head.  A 2 $1/2''$ speaker is mounted at the center {\it in} the head rather than {\it on} it.  In particular, the diaphragm of the speaker forms part of the pot outer wall.  Its motion compresses and expands the air inside, which is precisely the Helmholtz resonance motion.  In contrast, the $3/4''$ speakers mounted inside the sealed cylinders act as wigglers, producing both compression and expansion (at slightly different places) inside the cylinder.  (Both are pictured in FIG.~3.)
\begin{figure}[h!]
 \includegraphics[width=6.5in]{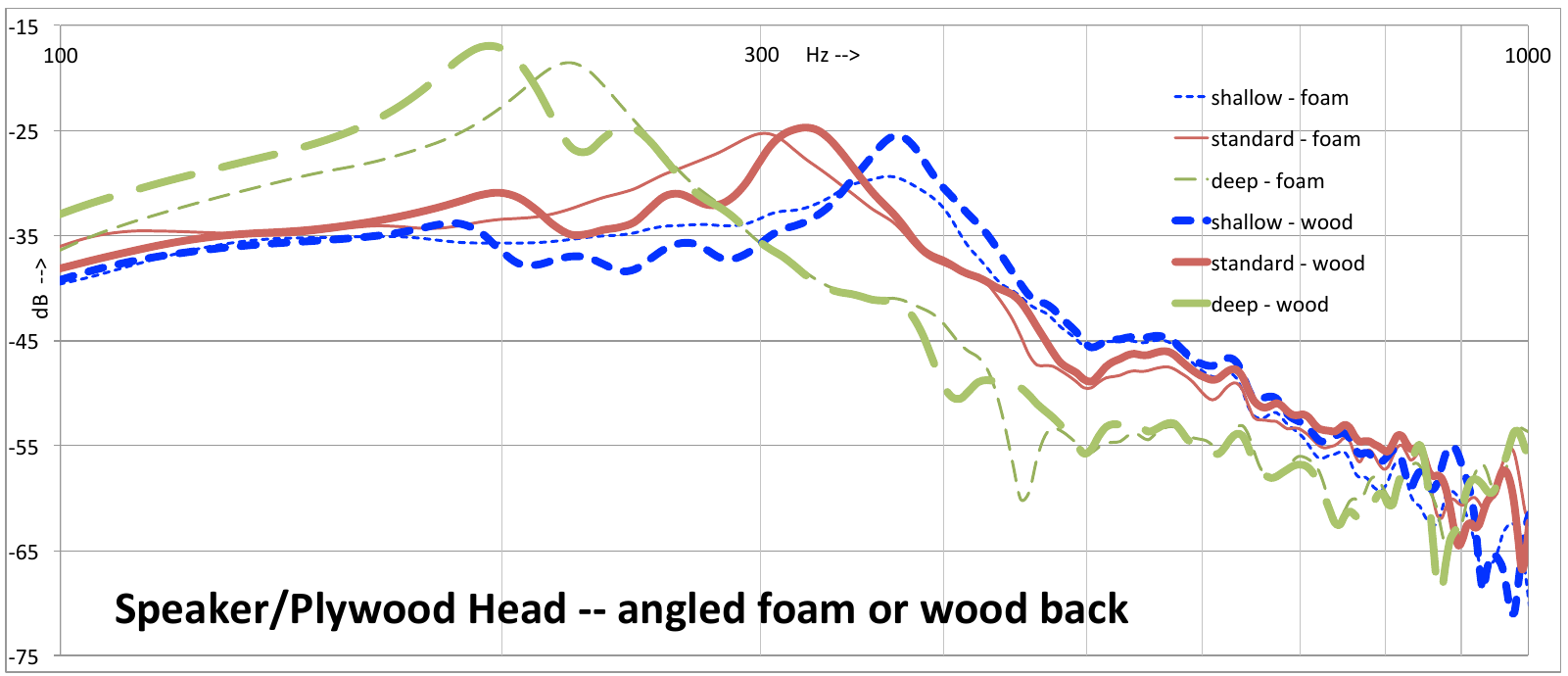}
 \caption{Helmholtz resonances excited by speaker in head with foam belly or wood back}
 \end{figure}
 
For these measurements, the pots have backs that simulate open-back playing.\cite{openback} In particular, there are runs with the foam-cork-Hawaiian shirt synthetic belly and runs with a plywood back.   Thin lines in FIG.~6 correspond to the foam back, and  thick lines are for the wood back.  For both backs, the sound hole was defined by a $3/8''$ spacer placed at one point between the back and the rim, with the back touching the rim diametrically opposite.  This arrangement was the closest reproducible set-up I found to natural open-back playing, where the player's body is the back.\cite{openback}  The microphone was placed at $2''$ from the sound-hole opening in the back.  The foam backs seem to smear out some of the detailed features that are present with the wood backs.  It is easy to imagine that the foam flexes a bit and absorbs, while the wood reflects better but might rattle.

Finally, I examine the combination of a standard head and a normal (albeit synthetic) back.  Again, a long series of center head taps with a piano hammer are recorded.  In this case, the back is the foam-cork-Hawaiian shirt combination, spaced and angled as described above.  
\begin{figure}[h!]
 \includegraphics[width=6.5in]{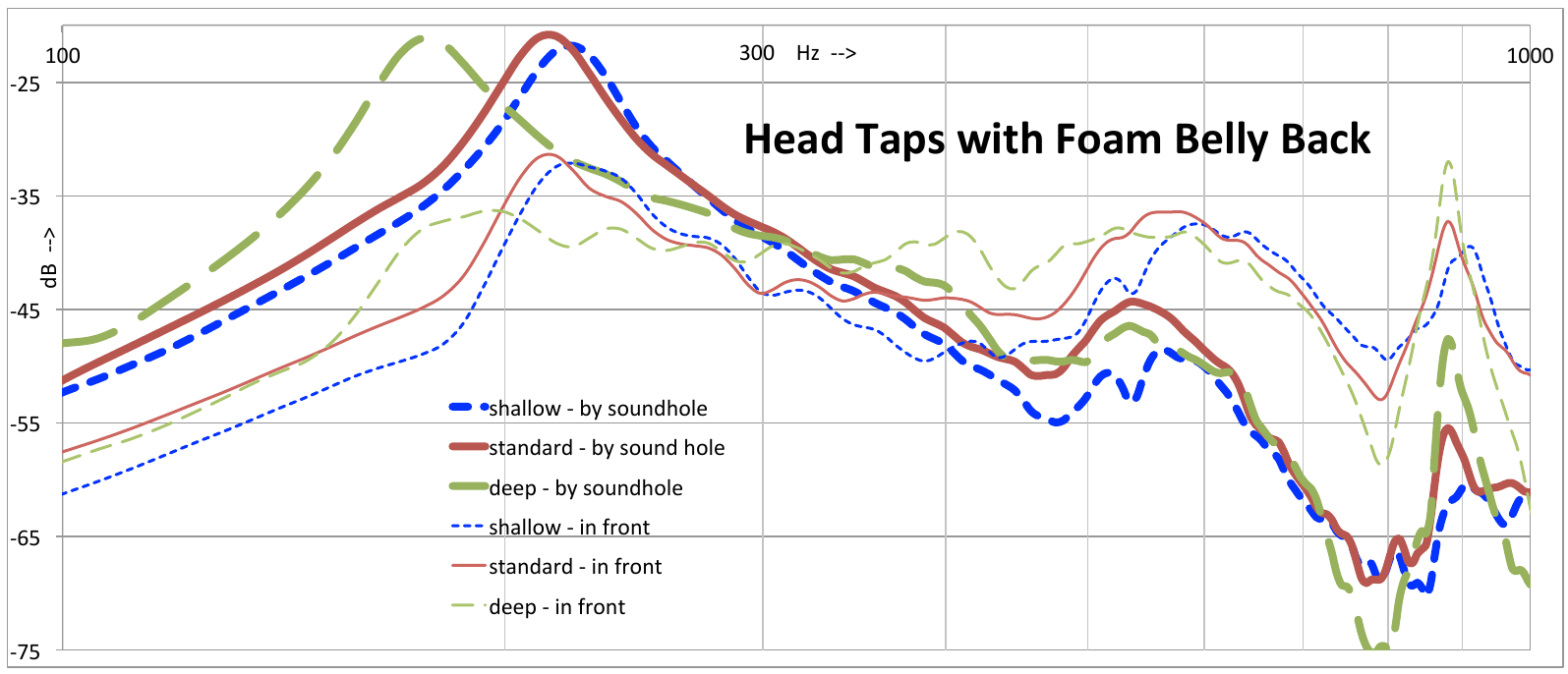}
 \caption{Head taps with the synthetic belly back and different mic positions}
 \end{figure}
\noindent For each of the three pots, FIG.~7 displays the spectra for two different microphone locations.  The thick lines are the result of mic placement at $2''$ from the sound hole in back.  That emphasizes the sound of the Helmholtz resonance and head motions that require a net movement of air in and out of the pot.  The thin lines are for mic placement at $20''$ in front of the head.  That is closer to how the banjo is played and heard.  The role of the in-and-out air motion is still quite evident,  but its magnitude in the signal is reduced relative to the other features of the sound.

FIG.~7 offers an example of how the physics works in these situations.  The idealized version of a drum tap on the head should be able to excite all resonances present --- at least to the extent that the tap is not near a node of that resonance.  One example of this stands out.  In addition to the low-lying resonances whose frequencies are rim-height dependent, there is clear evidence in FIG.~7 of a strong resonance which is nearly the same for all pots.  That's the one between 800 and 900 Hz.  The obvious interpretation is that these are the lowest ``closed" cylinder resonances.  As long as the taps were not all exactly at the center (which is on a pressure node line for those lowest cylinder resonances), they should be excited to some extent.  Not only are they rim-height independent as expected, they are clearly there when the pot is not sealed and when the head is allowed to vibrate.  And that is why the calculation and measurement of the sealed cylinders is not irrelevant.

\begin{figure}[h!]
 \includegraphics[width=4.5in]{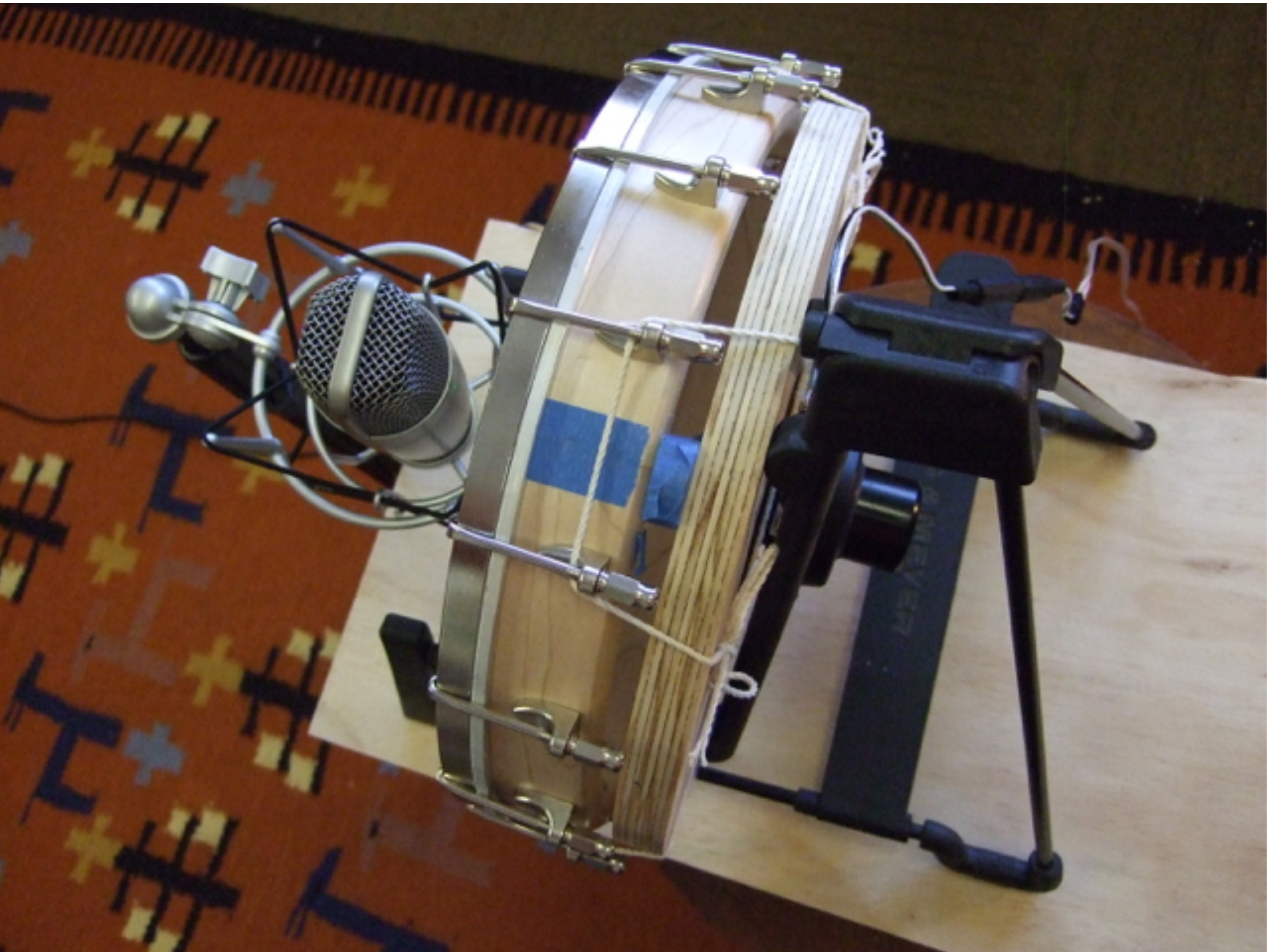}
 \caption{Set-up for one of the traces in FIG.~9: shallow pot; plywood/speaker back, angled with a $3/8''$ spacer; mic $2''$ from the center of the head}
 \end{figure}

FIG.~7 suggests the presence of modes between 500 and 600 Hz as well as those around 200.  So I tried yet another hardware approach to explore this region.  Classic studies of guitar acoustics offer a very satisfactory picture of the Helmholtz resonance mixing with the lowest sound board resonance to produce two distinct combinations.\cite{rossing} The spectrum has two resonances whose frequencies are functions of what would have been separate Helmholtz and sound board modes.  In a nice bit of elementary physics, when the two interact, they combine to give two distinct combinations with two new frequencies whose sum of squares is the same as the uncoupled case.

In an attempt to get a clearer picture of that frequency region, I tried the following, shown in FIG.~8.  I re-mounted the identical heads, tensioned again to 89 on a DrumDial, on the three rims.  For backs, I used the $3/4''$ plywood disks with 2 $1/2''$ speakers mounted in their centers.  The sound hole was defined by the $3/8''$ spacer at one point along the rim bottom, with the rim and back in contact at the diametrical opposite.  And I recorded in three different locations:  $2''$ from the sound hole spacer; in front of the head at $2''$ from the center; and at $20''$ from the head center.  (For the farther distance, the power to the speaker was increased by a factor of 100 to get a recorded sound of comparable strength as the others.)

\begin{figure}[h!]
 \includegraphics[width=6.5in]{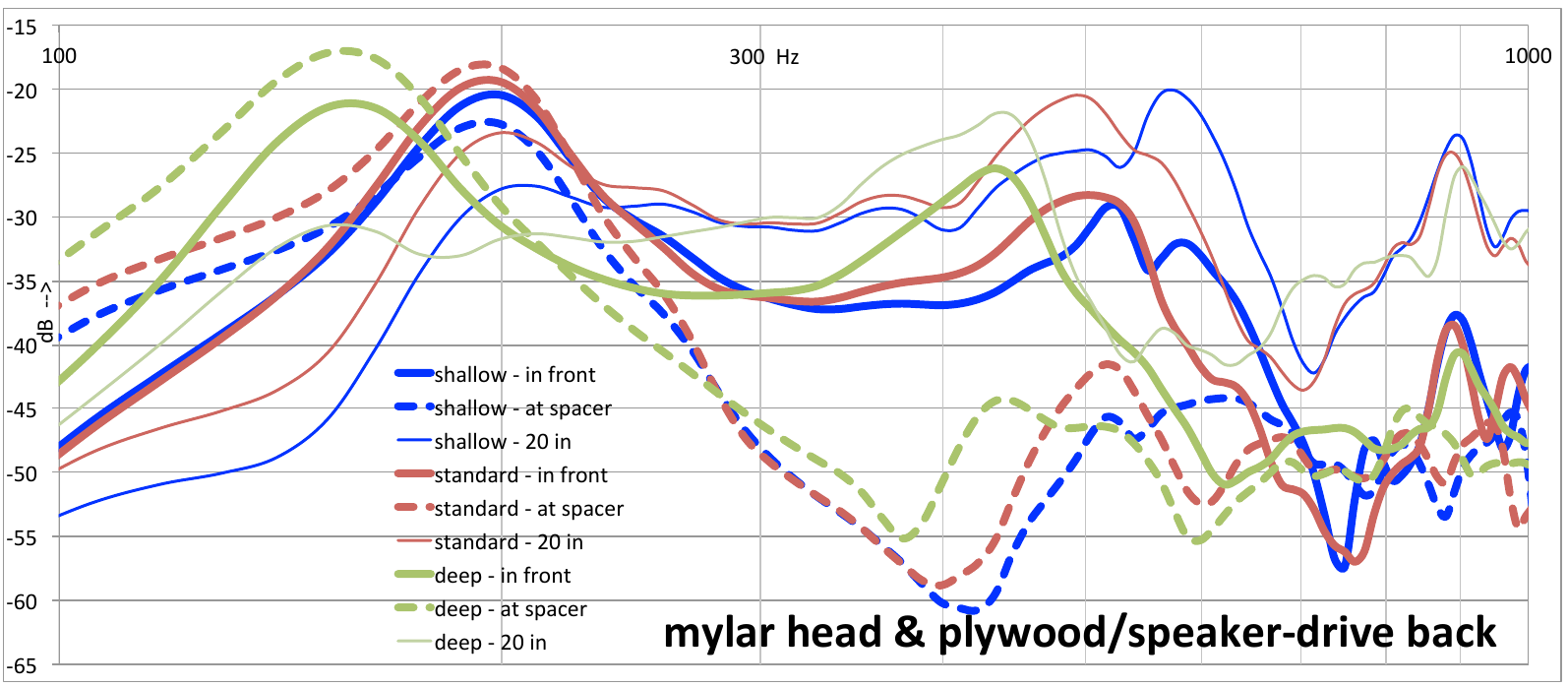}
 \caption{Response of the three rims to speaker in $3/4''$ plywood back, with mic in three positions}
 \end{figure}

The results are plotted in FIG.~9.  The clear similarity with FIG.~7 of the high amplitude features is actually evidence in support of the head-tap method.  From FIG.~9, in spite of all the apparent wiggling, I conclude that there are, indeed, three relevant resonances for each of the pots below 1000 Hz.  The lowest ones are between 150 and 200 Hz; the next are between 425 and 575 Hz; and the highest are around 900 Hz.  Theory suggests that these two lower resonances are both combinations of the Helmholtz and lowest head resonances.  They decrease in frequency with increasing rim height.  The highest is the same for all rims and is the lowest ``closed" cylinder mode.  The subtle details give additional support to these identifications.

The 900 Hz resonance appears with the microphone in front and not particularly when the mic is close ``at the spacer," i.e., at the sound hole.  That makes sense because the lowest closed-cylinder resonance involves air motion in the $r$-$\theta$ plane, moving from side to side.  The pressure is higher on one side than the other, alternating back and forth at $\sim$900 Hz. Those pressure variations push on the head and contribute to its up and down motion.  However, there's little reason for much air to venture out through the sound hole.  In contrast, the lowest head mode pushes air in and out the sound hole.  And similar air motion is a defining part of the Helmholtz resonance.  What happens typically\cite{rossing} is that the lower frequency combined motion occurs with both head and air volume pushing and pulling in the same direction at the same time.  In that case, there is air motion in and out of the sound hole and vibration of the enter of the head.  The typical higher frequency combined motion has the two effects opposing each other.  The head still moves and makes sound but the net motion at the sound hole is far smaller --- because the head is pushing it one way and the internal air is pushing it the other way.  In the measured sound, the lowest resonance is comparably visible in front and at the sound hole.  The second higher resonance is much stronger in front than at the sound hole.

Recording in a modest size room definitely produces wiggles in measurements of this sort.  The frequency of the speaker is swept very slowly through some range.  There is always a  room resonance very near to the driving frequency.  That sets up standing waves with nodal planes.  As the frequency slowly shifts, those planes move about.  In particular, they pass through the fixed location of the microphone.  So the  sound volume recorded at the mic goes up and down, even without any appreciable change in the whole-room average of the sound volume.  This was particularly noticeable when I took the recordings, standing a few feet from the microphone.  A display of the microphone voltage showed amplitude variations that were not particularly in sync with the variations I heard.  Also, a slight motion of my own head could make the sound dramatically louder or softer.  This would not happen in an anechoic chamber.  One can also eliminate these effects my recording outdoors in a big field with the microphone on the ground.  Neither of those were practical, available alternatives.

%\bigskip

One further check of this mixing interpretation for the two lowest resonances is a comparison of pots at different head tensions but the same rim height.  (I chose the deep rim because the wood speaker back did not have to be removed to change the head tension.)   I started with the carefully prepared 89 on the DrumDial by each hook, then did a run with all nuts tightened by $360^{\text{o}}$,  then one with all loosened by $180^{\text{o}}$ from the original 89 setting, 
\begin{figure}[h!]
 \includegraphics[width=6.5in]{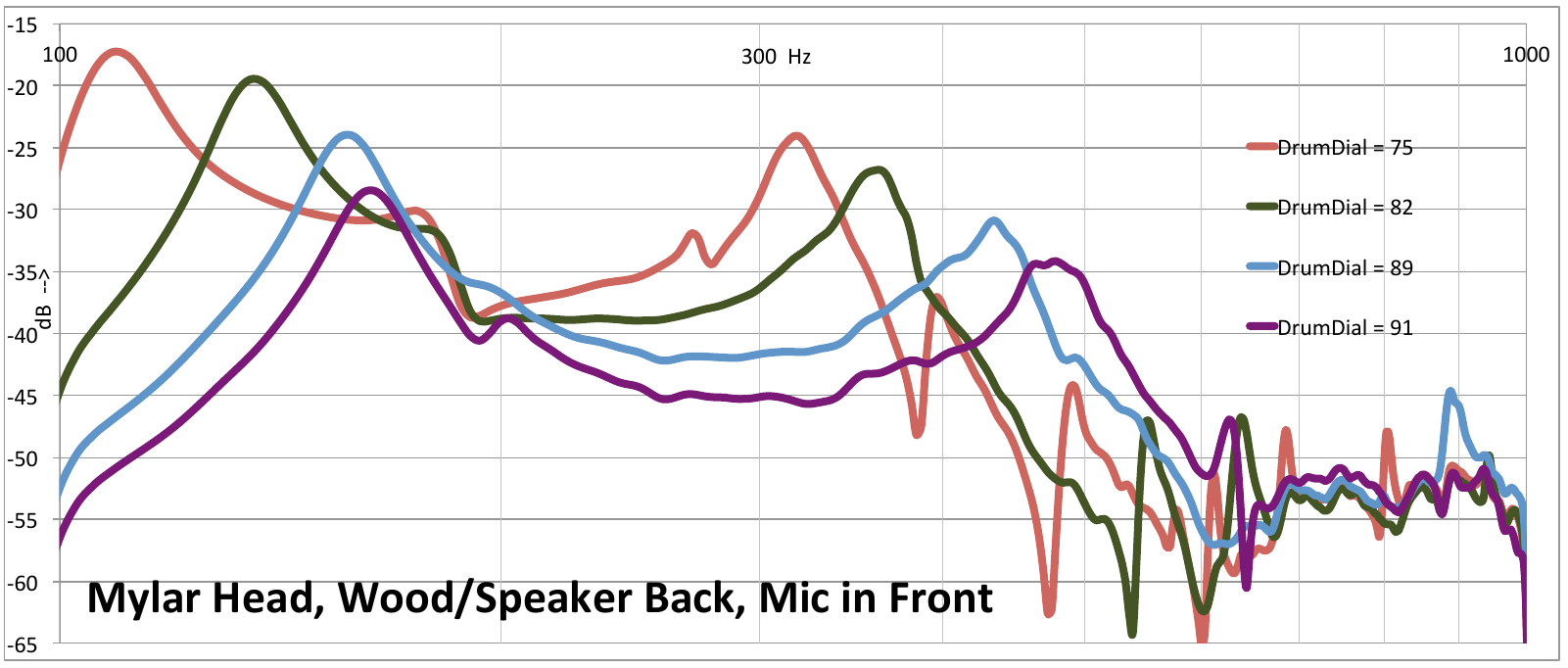}
 \caption{Different head tensions labeled by approx. DrumDial reading}
 \end{figure}
and then loosened another $180^{\text{o}}$.  I did not take care to even out the tensions at these new settings, but the approximate DrumDial readings were 91, the original 89, 82, and 75.  (91 is bright bluegrass tight, and any higher gets into the realm of ``tighten 'till it breaks and then back off a quarter turn.")  All tensions produced the same distinction between mic in front {\it versus} mic at the sound hole:  The lowest frequency peaks were of similar strength at both mic locations, while the second peaks, i.e., between 300 and 500Hz, were much weaker with mic at the sound hole than ain from of the head.  I only display the mic-in-front results in FIG.~10.  Also, the FIG.~10 frequency resolution is double that of FIG.~9.  That makes the positions of the highest peaks clearer, but it also reports some of the jitters that are artifacts of room sound.

The frequencies of the two lowest peaks increase with increased head tension.  That means head motion is significant for both of them.  On the other hand, there is a Helmholtz resonance in this region.  Taken by itself, its frequency is independent of the head tension.  So the simplest interpretation is that there is also only one head resonance in this region, but the actual resonant modes are (orthogonal, linear) combinations of the head and Helmholtz modes, with the resulting mic location dependence as discussed above.

In viewing and interpreting FIG.s~9 and 10, it's important to remember what's going on.  These banjos are systems with a great many parts.  The challenge is to understand important aspects of their performance by identifying a small number of crucial features and parts and developing a simple picture of how some idealization of those parts would behave.  Success is reckoned by how well a simple, understandable model represents and reproduces the observed features of the real systems.  At least, that's what interests me.

\section{why resonant modes and not ballistic propagation --- or waves and not particles}

Under normal circumstances, we do not see sound waves.  Nevertheless, people often imagine what's going on.  ``The sound goes here, bounces off, and then goes there," is sometimes said, as if describing a stream of bullets or rubber balls.  With somewhat greater sophistication, focusing, analogously to light and curved mirrors, is added to the description.  Indeed, in the case of light, lenses and mirrors can often be well-described by ray optics, in which a bundle of bullet-like ray trajectories are analyzed and combined.  There certainly are circumstances where sound behaves quite analogously --- but not in the case of the internal workings of musical instruments themselves.  For waves to act like particles and rays, their wavelengths have to be small compared to the other lengths of interest.  In optics, those much longer sizes might be the dimensions and curvature of the mirrors.  Musical sounds have wavelengths that start around $11'$ (around 100 Hz) and go to 1 $3/8''$ (around 10,000 Hz).  The physical features of musical instruments are not all much bigger, even for the highest pitches.

One basic aspect that distinguishes waves from particles is how they combine.  With two sources of particles, the particles add.  With two sources of waves, the waves combine so that at some places and times they might completely cancel, while at other places and times their combined intensity (loudness) is greater than the sum of the two.  There are ways to build up wave physics from trajectories, but those trajectories have to be combined with this addition/cancellation aspect strictly respected.  There is certainly legitimacy to the notion that sound bounces in and around an instrument.  However, there are always many bounces for a musical sound, even if it dies off relatively quickly.  In most optics situations, the light bounces just once off a given mirror.  (There are high tech and laboratory situations where multiple bounces are relevant.  In such cases, considering the entire light field all at once is usually more effective than trying to combine the successive bounces.)  There are acoustics situations where just one bounce is the dominant effect.  But the wave aspects emerge as relevant when combining waves, even with just a couple of overlapping waves.  Musical sounds almost invariably involve many reflections.  Strictly speaking, the perspective offered by resonances rests on an assumption of infinitely many reflections and assumes a steady-state situation.  There are subtle issues for which that is misleading\cite{coupled-damped}, but it is generally a good starting point.

\section{Conclusion}
Everything works the way it should.  To get to the produced sound, there are certainly a lot of details and quantitative connections that are far beyond simple physics.  But several pieces of the puzzle have been examined here, and they behave as expected.  Increasing the pot depth lowers the frequency of the lowest vibrations that the banjo can produce.  That was discussed in an earlier paper.\cite{openback}  The additional perspective offered here is how the pot depth effects the whole spectrum of response.  Missing is a detailed picture of how the air resonances talk back to the head.  That's really a good question.

\bigskip


\begin{thebibliography}{99}
\bibitem{Goodtime}I originally chose Deering Goodtime banjos for my acoustics investigations because 1) they are about as identical as wood objects can be, being a combination of CNC fabrication and high quality hand finishing; 2) they are quality instruments; and 3) they are relatively inexpensive.  When I fist approached Greg Deering, requesting some special items and perhaps a deal on the price, he immediately offered to provide me with whatever I needed.  He has been of great help ever since, including advice and fabrication.
\bibitem{heights}$2.80''$ is the standard Goodtime rim and typical of today's banjos.  No one plays a $2''$ rim.  They sound too thin.  I know of only two 5 $5/8''$ rim banjos in existence, and neither gets played regularly.  :(   The role of the deep pot in this study was to get a range of sound measurements that clearly tested the physics ideas.
\end{thebibliography}
\end{document}